\documentclass[conference]{IEEEtran}
\IEEEoverridecommandlockouts

\usepackage{cite}
\usepackage{hyperref}
\usepackage{amsmath,amssymb,amsfonts}
\usepackage{algorithm}
\usepackage{algpseudocode}
\usepackage{graphicx}
\usepackage{grffile} 
\usepackage{textcomp}
\usepackage[table, svgnames, dvipsnames]{xcolor}
\usepackage{verbatim}
\usepackage{braket}
\usepackage{subcaption}
\usepackage{placeins}
\usepackage{multirow}
\usepackage{booktabs}
\usepackage{makecell}
\usepackage{amsmath,amssymb}
\usepackage{tikz}
\usetikzlibrary{arrows.meta,positioning,fit,calc,quotes}
\definecolor{darkgreen}{RGB}{0,100,0}

\def\BibTeX{{\rm B\kern-.05em{\sc i\kern-.025em b}\kern-.08em
    T\kern-.1667em\lower.7ex\hbox{E}\kern-.125emX}}

\begin{document}

\title{ QuMod: Parallel Quantum Job Scheduling on Modular QPUs using Circuit Cutting
}
\author
{\IEEEauthorblockN{Vinooth Kulkarni}
\IEEEauthorblockA{\textit{Dept. of Computer and Data Sciences} \\
\textit{Case Western Reserve University}\\
Cleveland, OH, USA \\
vxk285@case.edu}
\and
\IEEEauthorblockN{Aaron Orenstein}
\IEEEauthorblockA{\textit{Dept. of Computer and Data Sciences} \\
\textit{Case Western Reserve University}\\
Cleveland, OH, USA \\
aao62@case.edu}
\and
\IEEEauthorblockN{Xinpeng Li}
\IEEEauthorblockA{\textit{Dept. of Computer and Data Sciences} \\
\textit{Case Western Reserve University}\\
Cleveland, OH, USA \\
xxl1337@case.edu}
\and
\IEEEauthorblockN{Shuai Xu}
\IEEEauthorblockA{\textit{Dept. of Computer and Data Sciences} \\
\textit{Case Western Reserve University}\\
Cleveland, OH, USA \\
sxx214@case.edu}
\and
\IEEEauthorblockN{Daniel Blankenberg}
\IEEEauthorblockA{\textit{Department of Molecular Medicine}\\
\textit {Cleveland Clinic Lerner College of Medicine} \\
\textit{Case Western Reserve University}\\
Cleveland, OH, USA \\
blanked2@ccf.org}
\and
\IEEEauthorblockN{Vipin Chaudhary}
\IEEEauthorblockA{\textit{Case Western Reserve University}\\
\textit{Dept. of Computer and Data Sciences} \\
Cleveland, OH, USA \\
vxc204@case.edu}
\thanks{This research was supported in part by NSF Awards 2216923 and 2117439.}
}

\maketitle

\begin{abstract}
The quantum computing community is increasingly positioning quantum processors as accelerators within classical HPC workflows, analogous to GPUs and TPUs. However, many real-world applications require scaling to hundreds or thousands of physical qubits to realize logical qubits via error correction. To reach these scales, hardware vendors employing diverse technologies---such as trapped ions, photonics, neutral atoms, and superconducting circuits---are moving beyond single, monolithic QPUs toward modular architectures connected via interconnects. For example, IonQ has proposed photonic links for scaling, while IBM has demonstrated a modular QPU architecture by classically linking two 127-qubit devices. Using dynamic circuits, Bell-pair-based teleportation, and circuit cutting, they have shown how to execute a large quantum circuit that cannot fit on a single QPU. As interest in quantum computing grows, cloud providers must ensure fair and efficient resource allocation for multiple users sharing such modular systems. Classical interconnection of QPUs introduces new scheduling challenges, particularly when multiple jobs execute in parallel. In this work, we develop a multi-programmable scheduler for modular quantum systems that jointly considers qubit mapping, parallel circuit execution, measurement synchronization across subcircuits, and teleportation operations between QPUs using dynamic circuits.  
\end{abstract}

\section{Introduction}
Quantum computing has rapidly become a central focus among emerging technologies, with industries racing to demonstrate progress across domains such as Secure long-distance communication, molecular simulations and protein folding in healthcare, combinatorial optimization, and Quantum machine learning and generative AI applications. The main driver is the growing set of problems that challenge existing classical techniques and the available computational resources. Quantum computers promise speedups for certain classes of classically hard problems by exploiting quantum-mechanical principles such as superposition and entanglement. Qubits, the basic units of quantum computation, are two-level quantum systems whose states are vectors in a two-dimensional Hilbert space acted on by unitary operations. An $n$-qubit register lives in a $2^n$-dimensional state space, so operating on superposition states allows algorithms to process structured information over an exponentially large space.  

Despite this potential, current quantum hardware remains in the noisy intermediate-scale quantum (NISQ) \cite{preskill2018quantum} regime. Qubits are realized in several physical platforms---including superconducting transmons, trapped ions, photonics, and neutral atoms---each with different trade-offs in fidelity, connectivity, and scalability. Because these technologies are still maturing, devices suffer from decoherence and gate errors arising from environmental noise, imperfect calibration, and control cross-talk. 
In practice, coherence times are limited to at most a few seconds, and many architectures (e.g., superconducting qubits) have sparse planar connectivity that makes long-range two-qubit operations more error-prone than in trapped-ion systems. Error correction further multiplies the physical qubit count, making it extremely challenging to scale monolithic devices with sufficiently low noise.

To overcome these scalability limits, recent work explores \emph{modular quantum computing}, where multiple quantum processing units (QPUs) are interconnected to act as a single larger machine. Two main forms of interconnect have been studied. First, \emph{quantum} interconnects, such as photonic links between trapped-ion chains, aim to distribute entanglement directly across modules \cite{Main_2025}. Second, \emph{classical} interconnects, such as IBM's real-time classical link between two 127-qubit Eagle QPUs \cite{Carrera_Vazquez_2024}, enable dynamic circuits in which operations on one QPU are conditioned on mid-circuit measurement outcomes from another, effectively realizing two-qubit interactions between qubits on different QPUs via teleportation-based circuit knitting \cite{qiskit-addon-cutting}.

Circuit cutting partitions a large circuit that would otherwise require more qubits than available on a single device by cutting wires or gates, producing multiple smaller subcircuits \cite{Tang_2021}. The partitioned subcircuits contain only local operations, replacing the gates (multi-qubit) that were cut. The expectation value of the original circuit is recovered via quasi-probability reconstruction based on the measurement outcomes of these subcircuits \cite{Piveteau_2024}.
However, the number of subcircuits grows exponentially with the non-local gates that need to be cut to partition a large circuit, which is termed as sampling overhead. There are mainly 2 ways of sampling the outcomes from the subcircuits. 1. Sampling based on local operations (LO), the cut qubits are treated purely classically: upstream subcircuits measure the cut qubits in an appropriate basis, downstream subcircuits are initialized in eigenstates, and classical post-processing (e.g., via Kronecker products and quasi-probability weights) reconstructs observables of the original circuit. 
The way a circuit is partitioned (wire-cuts versus gate-cuts) determines the number of subcircuits and the \emph{sampling overhead}: cutting $k$ wires under LO typically requires on the order of $16^k$ subcircuit sampling, and cutting a two-qubit gate (Example: controlled-NOT) can require up to nine subcircuits sampling. Various optimization techniques have been explored to reduce the sampling overhead in circuit cutting \cite{Chen_2023online,Perlin_2020,li2024efficient,chen2023efficient}. In all cases, the upstream subcircuits must be measured over appropriate Pauli bases, and the downstream subcircuits must be reinitialized accordingly to reconstruct the original bitstrings.


Sampling overhead drops significantly when the architecture supports \emph{local operations and classical communication} (LOCC) across QPUs: mid-circuit measurements on the upstream fragment are transmitted over a low-latency classical link to drive downstream conditional operations in real time using dynamic circuits (teleportation-style transfer of the cut qubit). Instead of treating the cut qubit as fully classical with offline post-processing, LOCC enables per-shot feed-forward control between fragments. For $k$ wire cuts, this reduces sampling overhead from $16^k$ to $4^k$ provided the feed-forward completes within coherence time, but it imposes architectural constraints requiring tightly synchronized mid-circuit measurements, classical links, and conditional gates.

Previous works in Quantum job scheduling are focused on running multiple circuits in parallel \cite{10.1145/3631525,10821381,kulkarni2025quflex}, accounting for measurement synchronicity and shot requirements, minimizing overhead associated with loading and unloading jobs, along with compilation and crosstalk errors due to parallel execution. Recent work on VQA scheduling~\cite{10764550,li2025qusplit} partitions training into exploratory and fine-tuning phases, assigning high-fidelity devices only to noise-sensitive (later) iterations while using lower-fidelity hardware for early, more noise-resilient steps, thereby improving overall fidelity under hardware constraints. To scale beyond the limitations of qubits when circuits are large and the exponential sampling overhead of $16^k$ subcircuits per wire cut, subcircuit scheduling with circuit-cutting techniques have been proposed \cite{kan2024scalablecircuitcuttingscheduling}. However, recent advancements in interconnects offer large circuits execution where the qubit requirement is beyond a single device. IBM has shown utilization of modular QPU architecture where a 142-qubit circuit is executed by cutting wires and teleporting the classical state after measurement. This approach reduces the sampling overhead exponentially from $16^k$ to $4^k$ for $k$ wire cuts.

In this work, we design a parallel quantum job scheduler on modular architectures with classical interconnects, taking into account the subcircuit sampling overhead and synchronicity required for subcircuit initialization based on dynamic circuits for measurement and feedback in two cases: (1) large circuits that cannot be run on a single device, where cutting is mandatory, and (2) dynamic decision making for circuit cutting in two modes (a) with Local operations (LO) (b) Local Operations with classical communication (LOCC) for maximum device utilization while taking into account device fidelity and the overall makespan of the job queue.

\section{Background}
\label{sec:background}

\subsection{Circuit cutting with local operations (LO)}
\label{subsec:lo_cutting}

 Consider a circuit $U$ acting on a register that we split into two disjoint subsets of qubits, $L$ and $R$, assigned to different devices.
 In the LO setting, each partition can only implement \emph{local} channels on the qubits it owns.  Any gate that acts jointly on qubits in $L$ and $R$ must therefore be removed and replaced by an arrangement of (i) completely local operations on $L$ and $R$ and (ii) classical post-processing of measurement outcomes.  This is achieved via a quasi-probability decomposition (QPD) of the non-local channel.  Concretely, if $G$ is a two-qubit gate that couples $L$ and $R$, we write its action on a density operator $\rho$ as
\begin{equation}
  \mathcal{G}(\rho)
  \;=\;
  G \rho G^\dagger
  \;=\;
  \sum_{\alpha=1}^{M} w_\alpha
  \bigl(\mathcal{L}_\alpha \otimes \mathcal{R}_\alpha\bigr)(\rho),
  \qquad
  \Lambda \equiv \sum_{\alpha=1}^M |w_\alpha|,
\end{equation}
where each $\mathcal{L}_\alpha$ and $\mathcal{R}_\alpha$ is a channel implementable using only local unitaries, measurements, and classical control on the $L$ and $R$ qubits, respectively, and $\{w_\alpha\}$ are real coefficients for explicit decompositions \cite{Piveteau_2024}.
In an LO simulation, each occurrence of $G$ is replaced by the following randomized procedure:
\begin{enumerate}
  \item Draw an index $\alpha$ with probability $p_\alpha = |w_\alpha|/\Lambda$.
  \item Apply the local channels $\mathcal{L}_\alpha$ on $L$ and $\mathcal{R}_\alpha$ on $R$ (which may include local measurements and classical control internal to each partition).
  \item At the end of the circuit, measure the observable of interest and multiply the raw outcome $o$ for that shot by the weight $\Lambda\,\mathrm{sgn}(w_\alpha)$.
\end{enumerate}
Averaging these reweighted outcomes over many shots produces an unbiased estimator of the target expectation value.  However, the variance is amplified by a factor $\Lambda^2$, so the number of circuit executions required to reach a given precision scales as $\mathcal{O}(\Lambda^2)$.  For typical two-qubit entangling gates (e.g., CNOT, CZ), optimal QPDs in the LO setting use a small number of local branches ($M\leq 9$) with $\Lambda^2 \approx 9$, meaning that each cut gate effectively multiplies the required samples by a constant factor.%

\medskip
\noindent\textbf{Wire cutting:}
Gate cutting targets specific two-qubit gates, whereas wire cutting breaks a qubit worldline at an intermediate time and treats the identity channel between two time slices as the object to be decomposed.  Let $q$ be a qubit that is cut between an ``upstream'' segment $U_{\mathrm{up}}$ and a ``downstream'' segment $U_{\mathrm{down}}$. If we view the identity
on $q$ as a two-qubit operator, it can be expanded in a Pauli basis as
\begin{equation}
  \mathrm{id}_q(\rho)
  \;=\;
  \sum_{P \in \{I,X,Y,Z\}} c_P\, P \rho P,
\end{equation}
with real coefficients $c_P$ satisfying $\sum_P |c_P| = \Lambda_{\text{wire}}$.
Operationally, this expansion is implemented by inserting, at the cut, a measurement-preparation scheme:
\begin{enumerate}
\item On the upstream subcircuit, qubit $q$ is measured in a basis associated
with a Pauli operator $P \in \{X, Y, Z\}$ (and, if required by the chosen
decomposition, in the computational basis). The pair consisting of the
measurement setting $P$ and the corresponding outcome $m \in \{+1, -1\}$
uniquely specifies the upstream branch of the decomposition.

  \item On the downstream subcircuit, the original qubit $q$ is discarded and
replaced by a freshly initialized qubit $q'$ prepared in the eigenstate
$\ket{\phi_{P,m}}$ of $P$ corresponding to the upstream measurement setting
$P$ and outcome $m$ (e.g., $\ket{\pm}$ for $X$, $\ket{\pm i}$ for $Y$, and
$\ket{0}, \ket{1}$ for $Z$). The downstream unitary $U_{\mathrm{down}}$ is
then applied to this reinitialized register.
\end{enumerate}
When this procedure is applied independently to $k$ cut wires, the resulting
estimator is a quasi-probability average over all combinations of upstream
measurement settings and downstream state preparations. In the LO setting
with Pauli bases, the optimal decomposition of a single wire has
$\Lambda_{\text{wire}} = 4$, implying a variance overhead of
$\Lambda_{\text{wire}}^2 = 16$ per cut wire and a total sampling cost that
scales as $\mathcal{O}(16^k)$. From the scheduler’s perspective, each cut
wire induces a family of upstream subcircuits measured in different bases
and a corresponding family of downstream subcircuits initialized in the
associated eigenstates. All nonlocal correlations between partitions are
reconstructed only \emph{offline} by classically reweighting these
measurement outcomes.

\subsection{Circuit cutting with local operations and classical communication (LOCC)}
\label{subsec:locc_cutting}
In the LOCC setting, the circuit is again partitioned across disjoint sets of qubits, but real-time classical communication between QPUs is now permitted. Mid-circuit measurements on one partition may condition subsequent operations on another, while each device continues to implement only local gates. At the channel level, a non-local operation is realized by a composition of local unitaries and measurements, classical transmission of the outcomes, and classically controlled corrections. The resulting classical branches induce a quasi-probability decomposition analogous to the LO case, but the availability of feed-forward admits constructions with a smaller $\ell_1$–norm and consequently reduced sampling overhead. For example, a remote entangling gate employing an entangled resource state, Bell measurements, and conditional corrections achieves a constant overhead scaling as $\mathcal{O}(4^n)$ for $n$ non-local gates, in contrast to the $\mathcal{O}(16^n)$ overhead of purely LO-based cutting.


\medskip
\noindent\textbf{Wire cutting with LOCC:}
Consider again a single-qubit worldline that is cut between an upstream
segment executed on device $L$ and a downstream segment executed on device
$R$. In contrast to the LO setting, where the identity channel on this wire
is replaced by a mixture of measurement–repreparation gadgets, LOCC
implements the same channel via a teleportation-style protocol. A typical
construction proceeds in three stages:

\begin{enumerate}
  \item \emph{Entangled resource preparation.}
  Devices $L$ and $R$ are supplied with a shared Bell pair
  $\ket{\Phi^+}_{a_L a_R} = \tfrac{1}{\sqrt{2}}(\ket{00} + \ket{11})$
  across ancillary qubits $a_L$ and $a_R$, prepared prior to execution of
  the cut circuit.

  \item \emph{Upstream execution and measurement:}  The data qubit $q$ on $L$ is evolved by the upstream subcircuit $U_{\mathrm{up}}$ up to the cut.  At the cut position, device $L$ performs a Bell-basis measurement on the pair $(q,a_L)$, e.g., by applying a fixed two-qubit Clifford circuit followed by computational-basis measurements.  This produces two classical bits $(m_Z,m_X)\in\{0,1\}^2$ that encode which Bell state was observed.
  \item \emph{Classical communication and downstream execution:}  The bits $(m_Z,m_X)$ are transmitted over the classical link to device $R$ within the coherence window.  Conditioned on these bits, $R$ applies single-qubit Pauli corrections $Z^{m_Z} X^{m_X}$ to the entangled partner $a_R$ and then continues the downstream subcircuit $U_{\mathrm{down}}$ on $a_R$ in place of the original qubit $q$.  After completing $U_{\mathrm{down}}$, $R$ measures the observable of interest on its local qubits.
\end{enumerate}
In the ideal, noise-free case, this protocol exactly reproduces the action of the identity channel from the cut position on $L$ to the start of $U_{\mathrm{down}}$ on $R$, so no additional quasi-probability reweighting is required beyond the averaging over the four possible Bell-measurement outcomes.  In the quasi-probability formalism, the different outcomes $(m_Z,m_X)$ and the corresponding conditional corrections define four classical branches whose probabilities are given by the underlying quantum mechanics, which leads to an effective $\ell_1$–norm squared of order $4$ per wire instead of $16$ as in LO.  When $k$ wires are cut and treated independently, the variance overhead of the estimator therefore scales as $\mathcal{O}(4^k)$, yielding an exponential improvement in sampling complexity over LO wire cutting.

From the scheduler's perspective, LOCC wire cutting turns each cut into a pair of causally ordered subcircuits: an \emph{upstream} fragment that ends with mid-circuit measurements on device $L$, and a \emph{downstream} fragment that begins on device $R$ only after the corresponding classical outcomes have been received and the appropriate corrections have been applied.  Upstream and downstream fragments can be parallelized across different jobs and devices, but for each logical cut the downstream fragment is constrained to start no earlier than the completion time of its upstream partner plus the classical-communication and control-latency budget.  All non-local correlations across the cut are now mediated on-line via the classical link and conditional gates, rather than reconstructed offline from independent measurement records as in the LO setting.

\subsection{Parallel circuit Scheduling challenges with LO vs LOCC}



Although LOCC-based circuit cutting can significantly reduce the sampling
overhead (e.g., from $16^n$ to approximately $4^n$ for $n$ cut wires),
it introduces several practical drawbacks compared to purely local
operations (LO). We summarize the main trade-offs in scheduling jobs in a quantum cloud with LO vs LOCC approach below.

\paragraph{Loss of fully independent parallel execution.}
Under LO, all fragments produced by circuit cutting are statistically
independent and can be executed in any order on any available QPU, with
no cross-fragment synchronization. In contrast, LOCC introduces explicit
dependencies: downstream fragments must wait for measurement outcomes
from upstream fragments. This converts a set of
independent tasks into a multi-stage pipeline with precedence constraints,
thus reducing scheduling flexibility.

\paragraph{ Additional latency on dependent operations across groups}
When circuits are scheduled for parallel execution, circuits are grouped based on the circuit depths. As the slowest circuit in the group or the circuit with the largest depth decides the finish time of the entire group for each shot.
LOCC requires mid-circuit measurements, 
transmission of measurement outcomes between QPUs, followed by
classically controlled operations and post-processing. Even if the total number of required
shots are reduced for LOCC, each effective shot incurs extra latency:
\begin{equation}
    \Delta T_{\text{LOCC}} \approx
        n_{\text{cuts}} \cdot \beta_{\text{comm}} \cdot \tau_{\text{link}},
\end{equation}
where $n_{\text{cuts}}$ is the number of cut wires, $\beta_{\text{comm}}$
is the average number of transmitted bits per cut per shot, and $\tau_{\text{link}}$
is the inter-QPU classical link latency. For shallow or low-shot circuits,
this additional per-shot delay can offset the advantages of reduced
sampling overhead.
\paragraph{More complex error and noise behavior.}
LO-based circuit cutting adds variance through quasi-probability
reconstruction but keeps a simple noise model: each fragment sees only
local device noise. LOCC introduces extra noise from mid-circuit
measurements, classical communication, and conditional operations driven
by imperfect measurement outcomes. While the theoretical $4^n$ scaling
assumes ideal classical links and logic, these additional error sources
degrade practical fidelity and complicate error analysis.

\medskip
We use the \texttt{qiskit-addon-cutting}~\cite{qiskit-addon-cutting} library
to obtain a baseline LO decomposition: given an input circuit $C_j$ and a
device-size constraint, it returns a cut circuit $\tilde{C}_j$ annotated
with cut two-qubit gates, the corresponding fragments $\{F_{j,k}\}$, the
number of cut gates $n^{\text{cut}}_j$, and an LO sampling-overhead estimate
$\kappa^2_{\text{LO}}(j)$. For the two-qubit cuts produced by
\texttt{qiskit-addon-cutting}, we observe nine fragments per cut on average,
and we therefore model the sampling overhead as
$\kappa^2_{\text{LO}}(j) \approx 9^{\,n^{\text{cut}}_j}$. QuMod uses the
same cut decomposition in both LO and LOCC modes: LO executes the fragments
with quasi-probability reconstruction under this overhead, whereas LOCC
reuses the cut locations but implements each non-local interaction via a
teleportation-style primitive with reduced effective sampling cost.

\medskip
\section{QuMod Scheduler}

We implement QuMod scheduling by extending the grouping and parallel scheduling strategy from~\cite{10821381,kulkarni2025quflex} to a modular, LO/LOCC-aware setting. We first group jobs using a dynamic programming algorithm adapted from~\cite{10.1109/71.210815}, but we now form \emph{upstream} and \emph{downstream} groups that respect the direction of classical communication between QPUs.

For each logical circuit that we cut, we index the cut by $k$ and denote its upstream and downstream subcircuits by $U_k$ and $D_k$, respectively. Every subcircuit job generated by this cut lies in either $U_k$ or $D_k$. We restrict attention to groups $g$ that are \emph{feasible}:

\begin{equation}
\mathcal{F}
= \Bigl\{ g \subseteq J \,\Big|\,
\sum_{j \in g} q_j \le Q_m,\;
g \cap U_k = \emptyset \,\lor\, g \cap D_k = \emptyset,\ \forall k
\Bigr\},
\label{eq:feasible_groups}
\end{equation}

i.e., the total qubit demand of $g$ must not exceed the capacity $Q_m$ of machine $m$, and for every cut $k$ a group may contain either upstream subcircuits ($g \cap U_k$) or downstream subcircuits ($g \cap D_k$), but never both. This prevents a single group from mixing causally dependent upstream and downstream subcircuits and preserves the LOCC execution order shown in Algorithm \ref{alg:qumod_grouping}.

In the grouping algorithm, we add the above constraints to the cost function,
\begin{align}
a(g) &= \frac{\max_{j \in g} T_j}{\min_{j \in g} T_j} - 1,
\label{eq:a_term} \\[2pt]
b(g) &= \lambda\bigl(\max_{j \in g} C_j - \min_{j \in g} C_j\bigr),
\label{eq:b_term}
\end{align}
where $a(g)$ measures runtime imbalance within the group and $b(g)$ penalizes its causal span. The overall cost is
\begin{equation}
d_{\text{qumod}}(g) =
\begin{cases}
\infty, & g \notin \mathcal{F}, \\
a(g) + b(g), & g \in \mathcal{F},
\end{cases}
\label{eq:partition_cost_qumod}
\end{equation}
so that infeasible groups are discarded, and the remaining groups are encouraged to be both runtime-balanced and causally tight.

Once groups are formed, each upstream and downstream group is mapped to machines based on the estimated makespan and fidelity of running that group in parallel. QuMod first generates an initial schedule Algorithm~\ref{alg:qumod_main} and then iteratively improves device utilization through adaptive circuit cutting. For each job or group of jobs assigned to a machine, QuMod estimates the qubit requirements and sampling overhead after cutting Algorithm~\ref{alg:qumod_cutting}. Candidate cuts that would exceed a global sampling budget (e.g., due to $9^k$ or $16^k$ scaling in the number of subcircuits) are discarded, so cutting is adaptive both to the available qubit slots and to the allowable sampling overhead.
\begin{figure}[htbp]
    \centering
    \caption{\small The figure shows the distributions of circuit characteristics of widths/depths for Random Heterogeneous Queue.}
    \label{fig:dist_jobs_queko}

  
    \begin{minipage}{0.95\columnwidth}
        \centering
        \includegraphics[width=\linewidth]{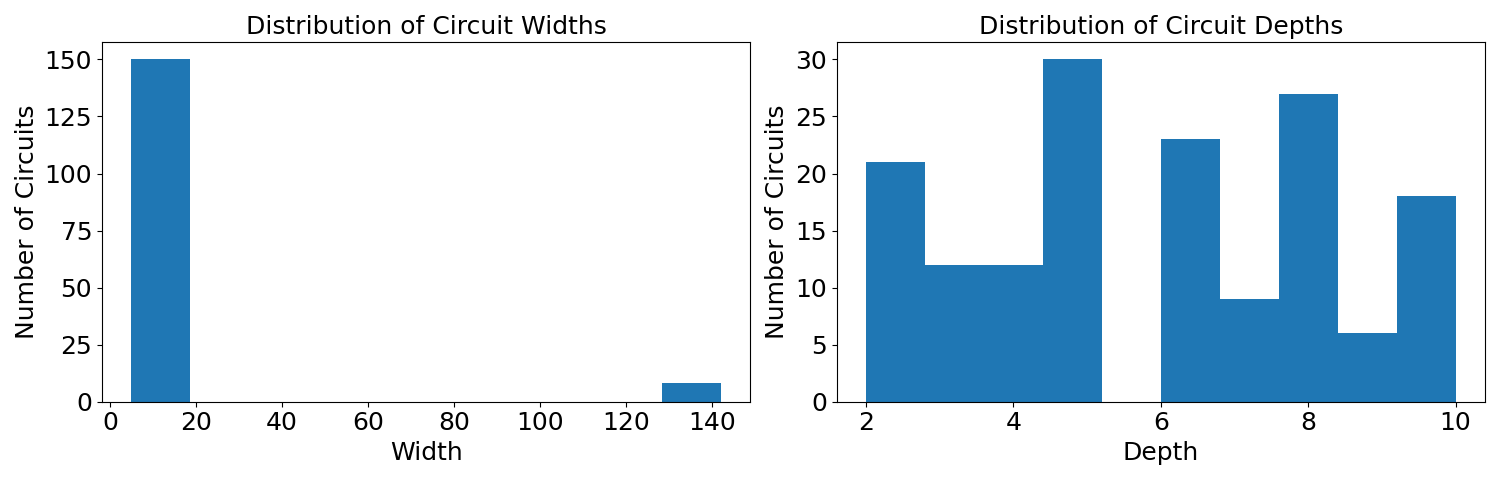}
    \end{minipage}
\end{figure}
To identify where additional subcircuits can be placed, QuMod scans the current multi-QPU schedule and computes the number of free ``slots'' across machines:
\begin{equation}
N_{\text{slots}} = \sum_{m \in \mathcal{M}} \left\lfloor \frac{Q_m - \sum_{j \in g_m} q_j}{Q_{\text{max\_sub}}} \right\rfloor,
\label{eq:utilization}
\end{equation}
where $g_m$ is the set of jobs currently assigned to machine $m$ and $Q_{\text{max\_sub}}$ is the maximum subcircuit size (in qubits) produced by the chosen cutting pattern from Algorithm~\ref{alg:qumod_main}. A cut is accepted only if the number of new subcircuits $J'$ can be placed in the available slots:
\begin{equation}
    |J'| \le N_{\text{slots}}.
    \label{eq:cut_condition}
\end{equation}

\begin{algorithm}[htbp]
\caption{QuMod Scheduling Algorithm (LO / LOCC Modes)}
\label{alg:qumod_main}
\begin{algorithmic}[1]
    \Statex \textbf{Input:} job queue $\mathcal{J}$, modular QPUs $\mathcal{M}$, $\mathtt{cut\_mode} \in \{\text{LO}, \text{LOCC}\}$
    \Statex \textbf{Output:} final schedule $S_{\mathrm{final}}$
    \State $\mathcal{J}_{\mathrm{current}} \gets \mathcal{J}$
    \Repeat
        \State $improved \gets \textbf{false}$
        \State $S_{\mathrm{initial}} \gets \textsc{GenerateInitialSchedule}(\mathcal{J}_{\mathrm{current}}, \mathcal{M})$
        \State $T_{\mathrm{initial}} \gets \textsc{Makespan}(S_{\mathrm{initial}})$
        \For{each job $j$ in $S_{\mathrm{initial}}$}
            \If{$j$ is not eligible for cutting}
                \State \textbf{continue}
            \EndIf
            \State $\text{sub\_jobs} \gets \textsc{TryCut}(j, \mathtt{cut\_mode})$
            \If{$\text{sub\_jobs} = \emptyset$}
                \State \textbf{continue}
            \EndIf
            \State $Q_{\max}^{(\mathrm{sub})} \gets \max_{j' \in \text{sub\_jobs}} q_{j'}$
            \State $N_{\mathrm{slots}} \gets 0$
            \ForAll{machine $m \in \mathcal{M}$}
                \ForAll{group $g$ scheduled on $m$ in $S_{\mathrm{initial}}$}
                    \State $Q_{\mathrm{used}} \gets \sum_{j' \in g} j'.\text{qubits}$
                    \State $Q_{\mathrm{avail}} \gets Q_{\mathrm{total}}(m) - Q_{\mathrm{used}}$
                    \State $N_{\mathrm{slots}} \mathrel{+}= \left\lfloor \frac{Q_{\mathrm{avail}}}{Q_{\mathrm{max\_sub}}} \right\rfloor$
                \EndFor
            \EndFor
            \If{$|\text{sub\_jobs}| > N_{\mathrm{slots}}$}
                \State \textbf{continue}
            \EndIf
            \State $\mathcal{J}_{\mathrm{cand}} \gets \big(\mathcal{J}_{\mathrm{current}} \setminus \{j\}\big) \cup \text{sub\_jobs}$
            \State $S_{\mathrm{cand}} \gets \textsc{GenerateInitialSchedule}(\mathcal{J}_{\mathrm{cand}}, \mathcal{M})$
            \State $T_{\mathrm{cand}} \gets \textsc{Makespan}(S_{\mathrm{cand}})$
            \If{$T_{\mathrm{cand}} \le T_{\mathrm{initial}}$}
                \State $\mathcal{J}_{\mathrm{current}} \gets \mathcal{J}_{\mathrm{cand}}$
                \State $improved \gets \textbf{true}$
                \State \textbf{break}
            \EndIf
        \EndFor
    \Until{not $improved$}
    \State $S_{\mathrm{final}} \gets \textsc{GenerateInitialSchedule}(\mathcal{J}_{\mathrm{current}}, \mathcal{M})$
    \State \Return $S_{\mathrm{final}}$
\end{algorithmic}
\end{algorithm}

In LO mode, QuMod treats all cuts as purely local: upstream and downstream subcircuits are scheduled on the same QPU with no inter-device communication. In LOCC mode, QuMod explicitly models classical communication and processing delays between upstream and downstream groups. For each cut that spans two QPUs, we insert a classical delay interval $\Delta_{\text{class}}$ consisting of measurement, transmission, and stitching time. Large circuits that generate many subcircuits incur larger $\Delta_{\text{class}}$, since more classical post-processing is required. These delays are inserted between the upstream and downstream groups and are taken into account when computing $N_{\text{slots}}$, so that downstream subcircuits only start after the corresponding classical data is available.

We end each iteration by reapplying the upstream/downstream grouping to close any scheduling gaps introduced by cutting and classical delays. If there are still jobs in the queue and free qubits across machines, QuMod repeats the cycle of grouping, adaptive cutting, and rescheduling. In this way, QuMod iteratively improves qubit utilization across modular QPUs while respecting both hardware constraints and sampling-overhead limits in LO and LOCC modes.

\begin{algorithm}[htbp]
\caption{QuMod Circuit Cutting Modes (LO / LOCC)}
\label{alg:qumod_cutting}
\begin{algorithmic}[1]
\Function{TryCut}{job $j$, $\mathtt{cut\_mode}$}
    \State $(\tilde{C}, \text{metadata}) \gets \textsc{FindCuts}(j.\text{circuit})$
    \If{$\mathtt{cut\_mode}=\text{LO}$}
        \State $\mathcal{P} \gets \{(\textsc{PartitionProblem}(\tilde{C}), \text{``flat''})\}$
    \Else
        \State $(\mathcal{C}_{\uparrow},\mathcal{C}_{\downarrow}) \gets \textsc{PartitionLOCC}(\tilde{C})$
        \State $\mathcal{P} \gets \{(\mathcal{C}_{\uparrow},\text{``upstream''}),(\mathcal{C}_{\downarrow},\text{``downstream''})\}$
    \EndIf
    \State $\text{sub\_jobs} \gets \emptyset$
    \ForAll{$(\mathcal{C}, s) \in \mathcal{P}$}
        \ForAll{subcircuit $C' \in \mathcal{C}$}
            \State create sub-job $j'$ with circuit $C'$
            \State $j'.\text{parent\_id} \gets j.\text{id}$
            \State $j'.\text{stage} \gets s$
            \State $j'.\text{arrival\_time} \gets j.\text{arrival\_time}$
            \State $j'.\text{shots} \gets j.\text{shots}$
            \State $\text{sub\_jobs} \gets \text{sub\_jobs} \cup \{j'\}$
        \EndFor
    \EndFor
    \State \Return $\text{sub\_jobs}$
\EndFunction
\end{algorithmic}
\end{algorithm}

\begin{algorithm}[htbp]
\caption{QuMod Grouping with LO / LOCC Constraints}
\label{alg:qumod_grouping}
\begin{algorithmic}[1]
    \Function{PartitionQumod}{$\mathcal{J}$}
        \Statex \textbf{Input:} runtime map $T$, qubit sizes $q$, device capacity
        $Q_{\text{dev}}$, max group size $C_{\max}$, parent map \texttt{parent},
        stage map \texttt{stage}
        \State $\mathcal{U} \gets \mathcal{J}$
        \State $\mathcal{G} \gets \emptyset$
        \While{$\mathcal{U} \neq \emptyset$}
            \State $g \gets \emptyset$
            \State $Q_{\text{used}} \gets 0$
            \State $\mathcal{U}' \gets \textsc{SortByRuntime}(\mathcal{U}, T)$
            \ForAll{$j \in \mathcal{U}'$}
                \If{$|g| = C_{\max}$}
                    \State \textbf{continue}
                \EndIf
                \If{$Q_{\text{used}} + q(j) > Q_{\text{dev}}$}
                    \State \textbf{continue}
                \EndIf
                \State $conflict \gets \textbf{false}$
                \ForAll{$k \in g$}
                    \State $p_j \gets parent(j)$
                    \State $p_k \gets parent(k)$
                    \State $s_j \gets stage(j)$
                    \State $s_k \gets stage(k)$
                    \If{$p_j \neq \bot \land p_j = p_k \land s_j \neq s_k$}
                        \State $conflict \gets \textbf{true}$
                        \State \textbf{break}
                    \EndIf
                \EndFor
                \If{$conflict$}
                    \State \textbf{continue}
                \EndIf
                \State $g \gets g \cup \{j\}$
                \State $Q_{\text{used}} \gets Q_{\text{used}} + q(j)$
            \EndFor
            \State $\mathcal{G} \gets \mathcal{G} \cup \{g\}$
            \State $\mathcal{U} \gets \mathcal{U} \setminus g$
        \EndWhile
        \State \Return $\mathcal{G}$
    \EndFunction
\end{algorithmic}
\end{algorithm}

\section{Evaluation and Observations}
\begin{table}[htbp]
    \centering
    \caption{\small QuMod LO vs LOCC for small, large, and random $>127$-qubit workloads.}
    \label{tab:qumod_lo_locc}
    \vspace{0.5ex}
    \begin{tabular}{|c|}
        \hline
        \\[-0.5ex]
        \begin{minipage}{0.95\columnwidth}
            \centering
            \begin{tabular}{|c|c|c|c|c|c|}
                \hline
                \textbf{Mode} & \textbf{Length} & \textbf{T\textsubscript{wait}} & \textbf{T\textsubscript{run}} & \textbf{T\textsubscript{total}} & \textbf{LPST} \\
                \hline
                QuMod LOCC 
                & 9.87  
                & 6.47  
                & 4.11  
                & 10.58 
                & \textbf{\textcolor{green!50!black}{-3.67}} \\ 
                \hline
                QuMod LO 
                & \textbf{\textcolor{green!50!black}{9.22}}  
                & \textbf{\textcolor{green!50!black}{6.24}}  
                & 4.11  
                & \textbf{\textcolor{green!50!black}{10.34}} 
                & -6.16 \\                                   
                \hline
            \end{tabular}
            \vspace{0.6ex}
            \caption*{\footnotesize
            \textbf{Small circuits (MQT-QUEKO).} 
            Workload changes: 32 (LOCC), 31 (LO).}
        \end{minipage}
        \\[1.2ex]
        \hline
        \\[-0.5ex]
        \begin{minipage}{0.95\columnwidth}
            \centering
            \begin{tabular}{|c|c|c|c|c|c|}
                \hline
                \textbf{Mode} & \textbf{Length} & \textbf{T\textsubscript{wait}} & \textbf{T\textsubscript{run}} & \textbf{T\textsubscript{total}} & \textbf{LPST} \\
                \hline
                QuMod LOCC 
                & \textbf{\textcolor{green!50!black}{2.73}}  
                & \textbf{\textcolor{green!50!black}{0.90}}  
                & \textbf{\textcolor{green!50!black}{2.88}}  
                & \textbf{\textcolor{green!50!black}{3.78}}  
                & \textbf{\textcolor{green!50!black}{-2.24}} \\ 
                \hline
                QuMod LO 
                & 3.15   
                & 1.96   
                & 3.64   
                & 5.60   
                & -6.47  \\ 
                \hline
            \end{tabular}
            \vspace{0.6ex}
            \caption*{\footnotesize
            \textbf{Large circuits (mandatory cut).}
            Workload changes: 16 (LOCC), 19 (LO).}
        \end{minipage}
        \\[1.2ex]
        \hline
        \\[-0.5ex]
        \begin{minipage}{0.95\columnwidth}
            \centering
            \begin{tabular}{|c|c|c|c|c|c|}
                \hline
                \textbf{Mode} & \textbf{Length} & \textbf{T\textsubscript{wait}} & \textbf{T\textsubscript{run}} & \textbf{T\textsubscript{total}} & \textbf{LPST} \\
                \hline
                QuMod LOCC 
                & \textbf{\textcolor{green!50!black}{75.44}} 
                & \textbf{\textcolor{green!50!black}{23.64}} 
                & \textbf{\textcolor{green!50!black}{2.96}}  
                & \textbf{\textcolor{green!50!black}{26.60}} 
                & \textbf{\textcolor{green!50!black}{-2.30}} \\ 
                \hline
                QuMod LO 
                & 81.08  
                & 29.23  
                & 3.69   
                & 32.92  
                & -3.72  \\ 
                \hline
            \end{tabular}
            \vspace{0.6ex}
            \caption*{\footnotesize
            \textbf{Random (158 heterogeneous circuits).}
            Workload changes: 150 (LOCC), 198 (LO).}
        \end{minipage}
        \\[1.2ex]
        \hline
        \begin{minipage}{0.95\columnwidth}
            \vspace{0.5ex}
            \footnotesize
            \textbf{Length}: Avg.\ queue length; 
            \textbf{T\textsubscript{wait}}: Avg.\ queue time; 
            \textbf{T\textsubscript{run}}: Avg.\ runtime; 
            \textbf{T\textsubscript{total}}: Avg.\ response time ($T_{\text{wait}} + T_{\text{run}}$); 
            \textbf{LPST}: log probability of successful trial (closer to 0 is better). 
            \textbf{Workload changes}: number of times the executing workload configuration changes.
        \end{minipage}
        \\
        \hline
    \end{tabular}
\end{table}

For QuMod, we use a SimPy-based discrete-event simulator~\cite{10.7717/peerj-cs.103} with (i) a Poisson job queue, (ii) a set of modular QPUs configured from calibration data of eleven IBM Quantum devices, and (iii) the QuMod scheduler implementing LO/LOCC-aware policies. The number of shots per circuit scales with its volume (width × depth), starting from 1,000 and increasing with a factor of 1.5, so larger, more noise-vulnerable circuits receive more shots. All experiments use a fixed scheduling window of 50 jobs.

We use the following performance measures for our evaluations, shown in Table \ref{tab:qumod_lo_locc}.
\begin{itemize}
    \item \textbf{Average Queue Length:} Time-averaged number of jobs in the queue.
    \item \textbf{Average Queue Time ($T_{wait}$):} Average time a job spends waiting before execution.
    \item \textbf{Average Runtime ($T_{run}$):} Average time a job spends executing on a quantum device (including overhead).
    \item \textbf{Average Response Time ($T_{total}$):} Total time from submission to completion, $T_{wait} + T_{run}$.
    \item \textbf{Log Probability of Successful Trial (LPST):} 
    Logarithmic estimate of a circuit's success probability \cite{10.1145/3631525}. Values closer to $0$ indicate a higher expected probability of correct, error-free execution, serving as a fidelity-like reliability proxy under the backend error model.
\end{itemize}

\subsection{MQT Benchmark circuits: Smaller Circuits}
\begin{figure*}[hp]
    \centering
    \subfloat[QuMod LO mode\label{fig:qumod_lo}]{%
        \includegraphics[width=0.48\linewidth]{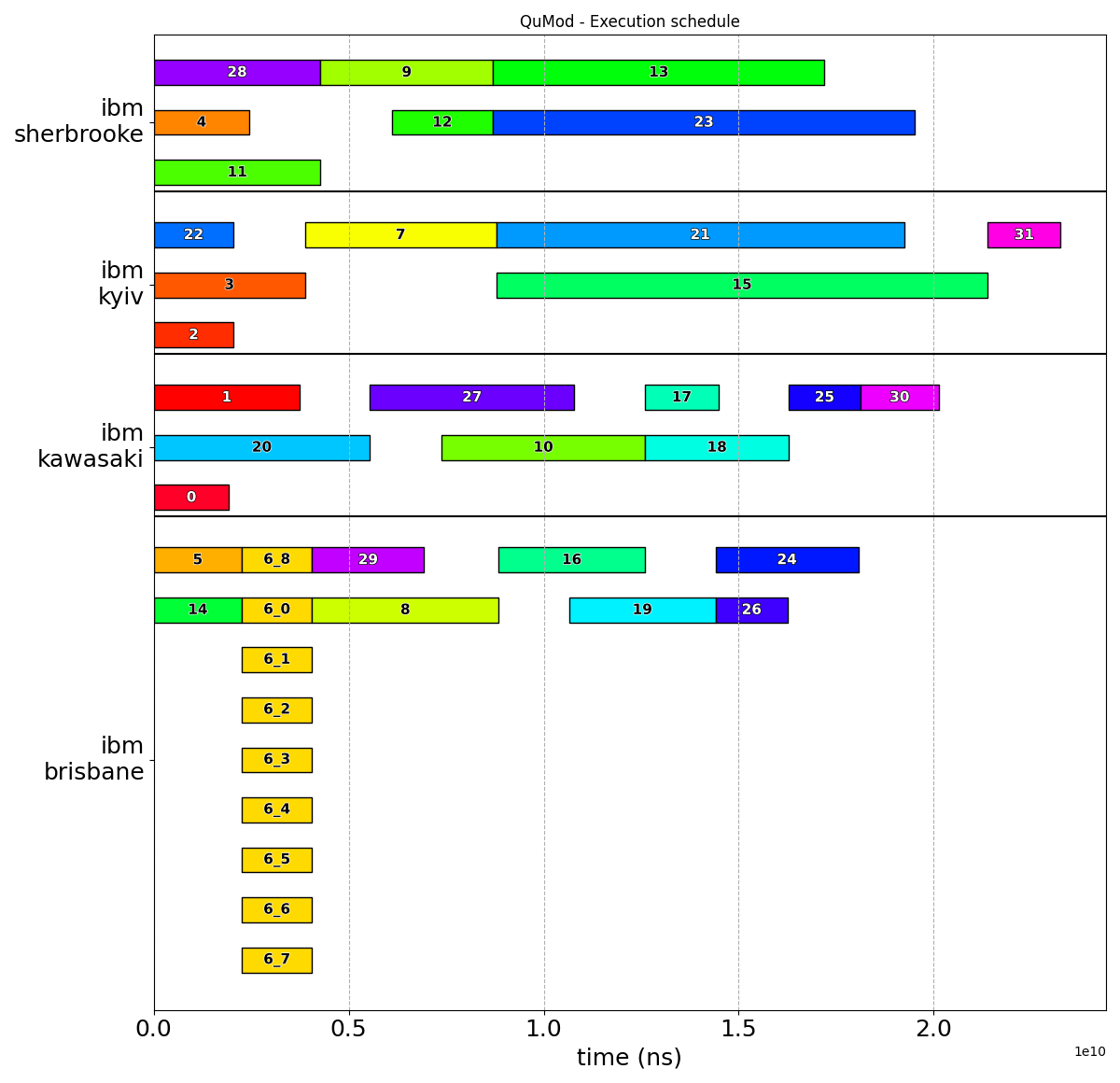}%
    }\hfill
    \subfloat[QuMod LOCC mode\label{fig:qumod_locc}]{%
        \includegraphics[width=0.48\linewidth]{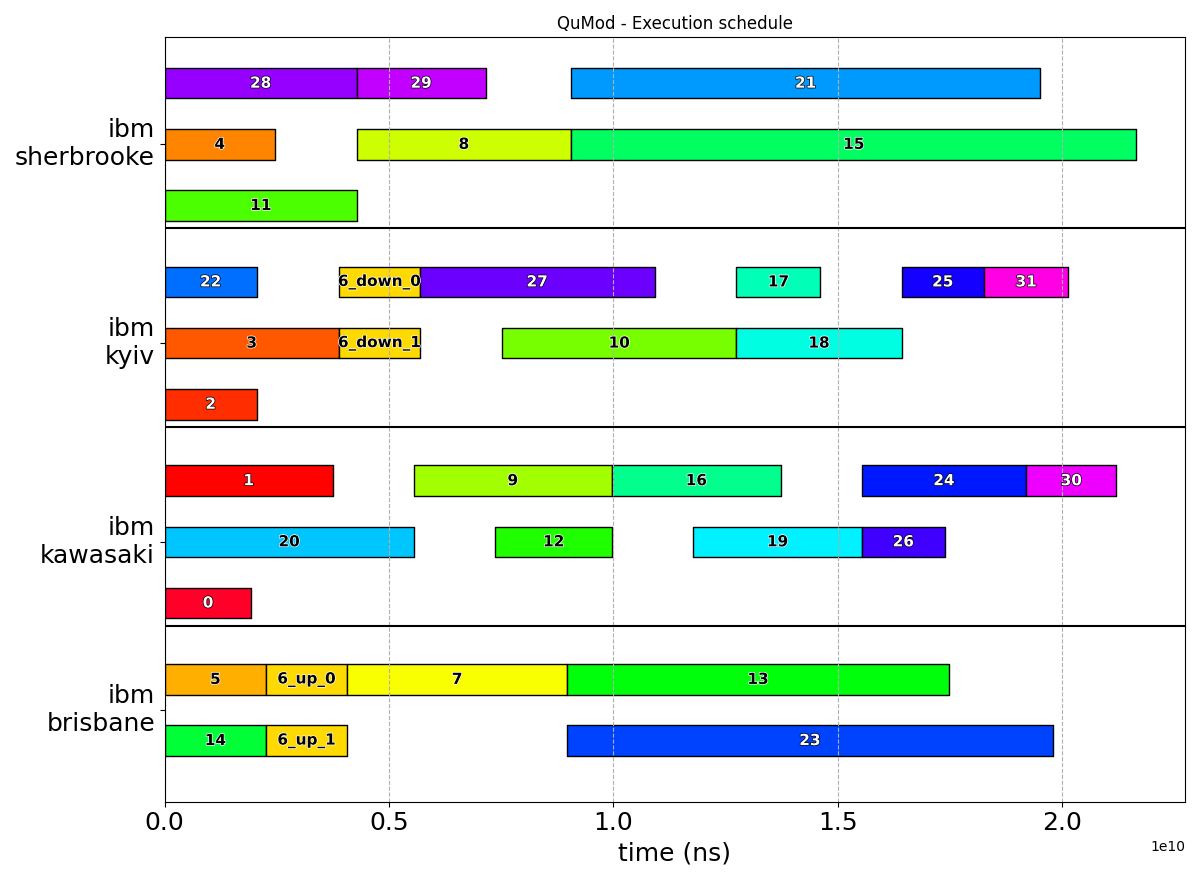}%
    }
   \caption{\small MQT-QUEKO Benchmark circuits: Execution schedules using LO and LOCC modes on modular QPUs. Black horizontal lines separate execution on each quantum computer. The numbering and coloring of jobs is consistent across both subfigures. (a) Subcircuits that can be run independently using circuit cutting with only local operations (LO). (b) Upstream subcircuits are scheduled before downstream subcircuits on another QPU connected via a classical communication link (LOCC).}
    \label{fig:lo_locc_mqt_queko}
\end{figure*}

In this setup, we run a queue of smaller MQT-QUEKO \cite{quetschlich2023mqtbench}circuits. We observe that the overall makespan is similar in both LO Figure  \ref{fig:qumod_lo} and Figure LOCC modes \ref{fig:qumod_locc}. In LO mode, the subcircuits can be executed independently; because they have similar depths/runtimes, they are grouped and run in parallel with high device utilization, as shown in Fig.~\ref{fig:lo_locc_mqt_queko}(a). 

In LOCC mode, the upstream subcircuits are grouped and executed on \texttt{ibm\_brisbane}, while the downstream subcircuits are run on \texttt{ibm\_kyiv}. The scheduler selects these devices based on the available backends at that point in time and chooses the configuration that yields the best estimated fidelity.

\begin{figure*}[hp]
    \centering
    \subfloat[QuMod LO mode\label{fig:142_qumod_lo}]{%
        \includegraphics[width=0.48\linewidth]{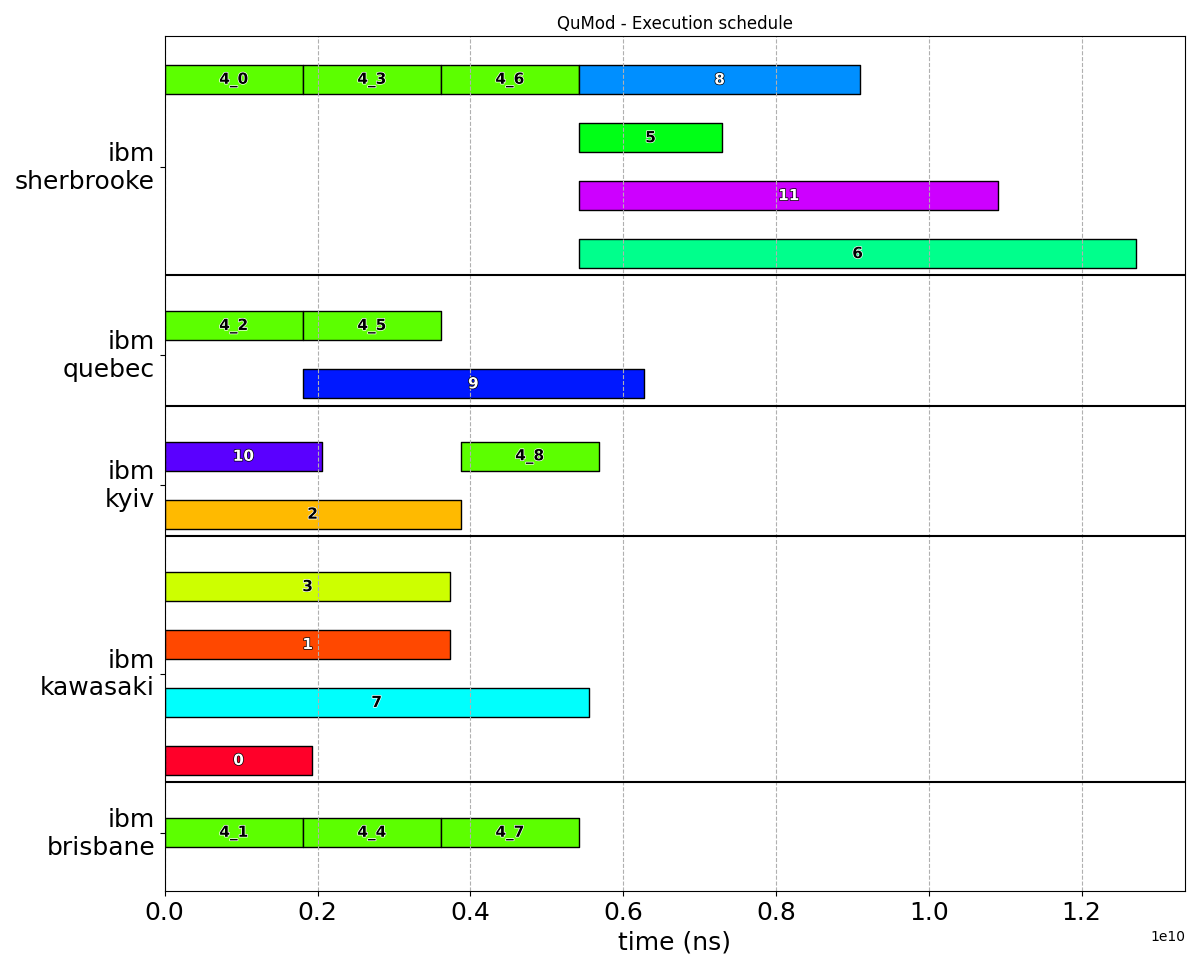}%
    }\hfill
    \subfloat[QuMod LOCC mode\label{fig:142_qumod_locc}]{%
        \includegraphics[width=0.48\linewidth]{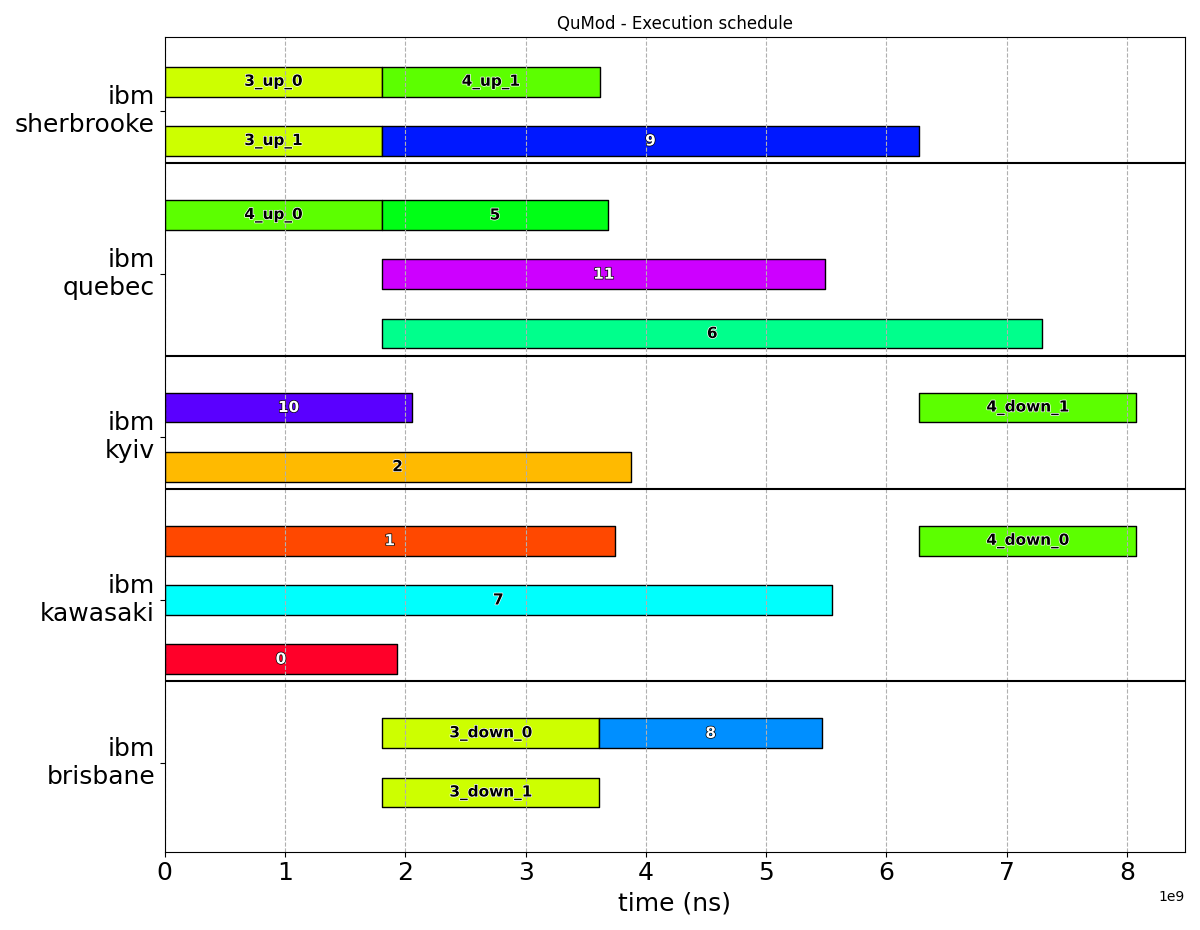}%
    }
 \caption{\small MQT + Large circuits (Mandatory cut): Execution schedules using LO and LOCC modes on modular QPUs. Black horizontal lines separate execution on each quantum computer. The numbering and coloring of jobs is consistent across both subfigures. (a) Only job 4 with 142 qubits was cut, which does not fit on a QPU. Cutting more circuits in LO mode produces more subcircuits which the scheduler avoided (b) QuMod dynamically selects additional job (job id = 3) that could be cut and scheduled for better qubit utilization across the QPUs.}
    \label{fig:142_lo_locc_mqt_queko}
\end{figure*}

\subsection{Large Circuits $>$ 127 Qubits}
In this setup, we combine smaller circuits with large circuits requiring more than 127 qubits, and evaluate the average fidelity and makespan of all jobs in the queue. Figure~\ref{fig:142_lo_locc_mqt_queko} shows the corresponding execution schedule. When the scheduler encounters a job requiring 142 qubits, which cannot be executed on any single device, that job is selected for cutting in both LO and LOCC modes.

In LO mode, the number of subcircuits scales as $9^k$ when each two-qubit gate (e.g., CNOT, CZ) is cut. Because the resulting subcircuits are still large (approximately 71 qubits), they cannot be easily grouped with other jobs for parallel execution, nor even with each other: running two such subcircuits in parallel would again require 142 qubits. As a result, all subcircuits must be executed largely sequentially, leading to a longer overall makespan shown in Figure \ref{fig:142_qumod_lo}.

In LOCC mode, shown in Figure \ref{fig:142_qumod_locc}, the scheduler additionally selects another circuit (job ID = 3) for cutting, based on device utilization. The resulting subcircuits are smaller and can be grouped into upstream and downstream sets that fit concurrently across the modular QPUs. This enables greater parallelism and improves overall qubit utilization.

\subsection{Random Heterogeneous Circuits}
We evaluate a mixed workload of large $(>127$-qubit$)$ and small circuits, with
the distribution shown in Fig.~\ref{fig:dist_jobs_queko}. Since none of the
large circuits fit on a single device, QuMod cuts them in both LO and LOCC
modes and groups the resulting subcircuits with small circuits of similar
depth for parallel execution. Across this workload, LOCC achieves higher
fidelity and lower response time than LO, as summarized in
Table~\ref{tab:qumod_lo_locc}.

\FloatBarrier

\section{Conclusion}
In this work, we introduced QuMod, a quantum job scheduler for modular QPUs that runs circuits in parallel and selectively applies circuit cutting in two modes (LO and LOCC) to improve qubit utilization in a cloud setting. QuMod explicitly accounts for classical communication between QPUs, synchronizing upstream and downstream subcircuits and grouping them with other jobs by runtime to enable parallel execution.
Because circuit cutting incurs exponential sampling overhead, naïve LO-style cutting quickly becomes impractical for large circuits with high qubit counts (e.g., when accounting for error correction). Our results show that the LOCC mode achieves better makespan and higher effective utilization by adaptively cutting and distributing circuits across QPUs while respecting sampling budgets. We evaluate QuMod using a SimPy-based simulator parameterized with IBM’s modular QPU architecture and classical interconnects, demonstrating that LOCC-aware scheduling can significantly reduce runtime overhead compared to purely local cutting while maintaining or improving fidelity.

\bibliographystyle{IEEEtran}
\bibliography{ref}

\end{document}